\documentstyle[aps,prl,preprint,epsfig,scref]{revtex} 
\draft
\begin{document}
\pagestyle{myheadings}
\markboth{Helbing/Treiber: Recent Breakthroughs in Traffic Theory}
{Helbing/Treiber: Recent Breakthroughs in Traffic Theory}

\title{Jams, Waves, and Clusters}

\author{Dirk Helbing\protect\footnotemark[1]
and Martin Treiber\\\mbox{ }} 
\renewcommand{\thefootnote}{\fnsymbol{footnote}}
\maketitle
\footnotetext[1]{II. Institute of 
         Theoretical Physics, University of Stuttgart,
         Pfaffenwaldring 57/III, 70550 Stuttgart, Germany,
         http://www.theo2.physik.uni-stuttgart.de/helbing.html}

Have you been suffering from traffic jams lately and asking yourself
why freeways are no free ways anymore?
After several great advances in traffic theory,
Korean physicists \cite{Lee} have now offered an interpretation of a 
recently discovered state of congested traffic, called ``synchronized''
traffic \cite{KR}. Their fluid-dynamic simulations can be a useful tool for
an optimization of traffic flow on motorways. 
\par
When Nagel and Schreckenberg presented their cellular automaton model
of traffic flow in 1992 \cite{NS}, 
allowing for a more than real-time simulation
of the entire road system of large cities, 
they probably did not anticipate the
flood of publications and the enthusiasm among scientists that
they would cause on the subject of traffic theory.
By treating huge numbers of interacting vehicles similar to classical
many-particle systems, physicists have recently added a 
lot to a better understanding of traffic flow. The
mathematical tools that they use, stemming mainly from statistical physics
and non-linear dynamics, have proved their interdisciplinary
value many times. This includes concepts reaching from 
self-organized criticality and phase transitions up to the
kinetic theory of gases, fluids, and granular media. 
\par
In traffic, drivers try
to maximize their own utilities (i.e. velocity, safety, and comfort)
within the constraints imposed by physical limitations and traffic
rules. Under certain conditions, their competitive, non-linear interactions 
give rise to the formation of collective patterns of motion like
traffic jams. 
The various observed phenomena on freeways are surprisingly rich:
Apart from free traffic and extended traffic jams behind bottlenecks,
there are localized clusters (small moving jams) and stop-and-go
waves. In addition, Kerner and Rehborn \cite{KR} have recently
discovered a hysteretic phase transition from free traffic to a
form of congested traffic (mostly appearing close to on-ramps)
that had not been identified in more than 40 years of traffic research,
they say. Kerner and Rehborn
call it ``synchronized'' traffic because of the synchronization
of velocities among lanes. However, 
the probably more characteristic feature
is its high flow in spite of the breakdown of velocity, 
which is in contrast to traffic jams. Downstream of the ramp,
the breakdown of velocity 
eventually relaxes to free traffic in the course of the freeway.
Another interesting property is
the wide scattering of synchronized traffic states, when plotted in
the flow-density plane, which differs from the quasi-linear 
density-dependence of free traffic flow.
\par
Lee {\em et al.} \cite{Lee} have now suggested an explanation for this
hysteretic phase transition. They simulated freeways, including
on- and off-ramps, 
with a fluid-dynamic traffic model that is closely
related to the Navier-Stokes equations for viscous, compressible fluids.
However, it contains an additional relaxation term describing adaptation of
average vehicle velocity to an 
equilibrium velocity, which monotonically decreases with growing density. 
In comparison with previous simulation studies, Lee {\em et al.} 
used another velocity-density relation and a considerably different set 
of parameters. 
By a temporary peak in the on-ramp flow, they managed to trigger a
form of oscillating congested traffic that is propagating upstream, but
pinned at the location of the ramp (see Fig.~1a). They call it the
``recurring hump'' state (RH) and compare it to autocatalytic
oscillators. Free traffic would correspond to a
point attractor and the oscillating traffic state to a stable limit cycle. 
In terms of non-linear dynamics, the transition corresponds to a 
Hopf bifurcation, but a subcritical one, since the critical
ramp flow depends on the size of the perturbation.
\par
Lee {\em et al.} point out that free traffic (FT) survives the assumed
pulse-type perturbation of finite amplitude, if the ramp flow is 
below a certain critical value. However, once a RH state has formed,
it is self-maintained until
the ramp flow falls below another critical value which is smaller
than the one for the transition from FT to the RH state. This proves
the hysteretic nature of the transition. Moreover, Lee {\em et al.}
could show the gradual spatial transition from the RH state to
free flow downstream of the ramp. They also managed to reproduce the
synchronization among neighboring freeway
lanes as a result of lane changes. Therefore, they suggest 
that their model can describe the empirically observed first-order phase 
transition to synchronized traffic. The two-dimensional scattering
of synchronized traffic states is understood as a result of 
the fact that the amplitude of the oscillating traffic state
depends on the ramp flow.
\par
Although the interpretation of synchronized traffic by Lee {\em et al.}
does not {\em quantitatively} agree with the observations,
in various respects it comes pretty close to reality. Meanwhile,
a more complete explanation has been offered \cite{HT}. 
Above all, the findings are also of great practical importance.
A more detailled analysis shows that there is a 
whole spectrum of different states that can form at ramps. 
Their occurence decisively depends on the inflow as well as the 
ramp length (see Fig. 1b). This is not only
relevant for an appropriate dimensioning of ramps, but also for an 
optimal on-ramp control.
\par
In conclusion, traffic theory is presently more interesting than ever before.
Recent advances have 
yielded a better understanding of traffic flow phenomena as well
as realistic and fast simulation models. Scientists are now prepared
to design on-line controls for efficient
traffic optimization, calculate the environmental impact of
congestion, and develop methods for traffic forecasts.\\[6mm]

\begin{figure}
\begin{center}
\hspace*{3\unitlength}\includegraphics[width=80\unitlength]
 {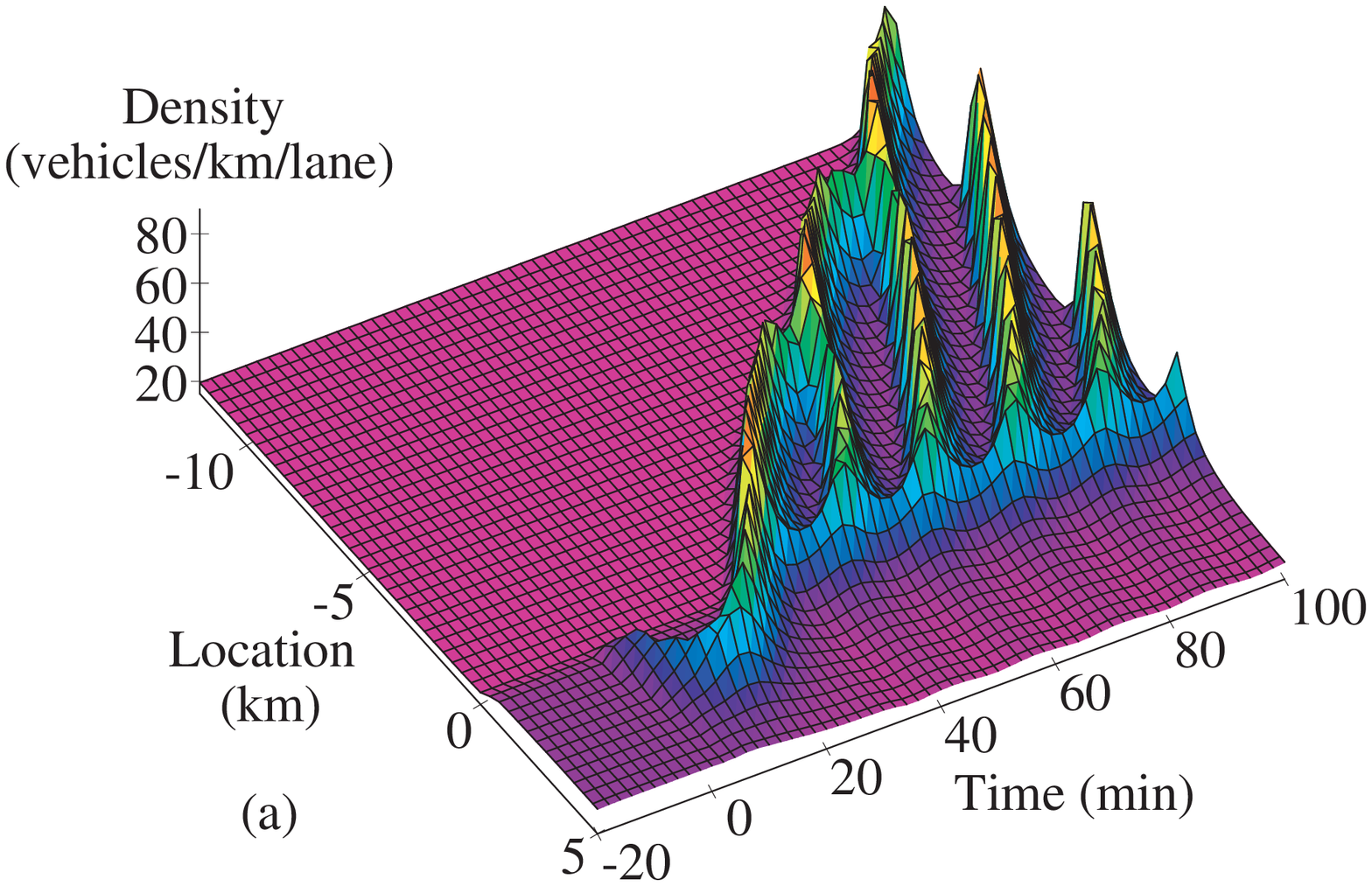} \\
\includegraphics[width=80\unitlength]
 {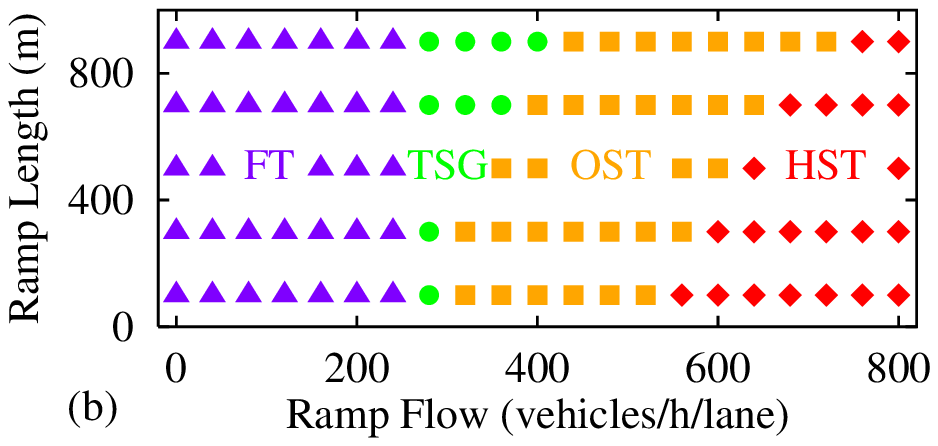}
\end{center}
\caption[]{(a) Formation of the recurring hump state (RH) on a freeway
for the model and parameters used by Lee {\em et al.} Since we used free
rather than periodic boundary conditions, the inflow at the on-ramp 
located at 0\,km did not need to be balanced
by an off-ramp, which is more realistic. The constant ramp flow
of 318 vehicles per hour and lane
causes a higher vehicle density downstream of the
ramp. Nevertheless, free flow pertains, 
until a short perturbation of the
ramp flow occurs at time 0\,min, which moves downstream in the beginning.
However, with growing amplitude, the perturbation
changes its propagation speed, reverses its direction, 
and finally induces another, bigger perturbation,
when passing the ramp. This process repeats again and again, in this
way generating the oscillating
RH state. When passing the ramp, the perturbations
continue their way upstream, 
until they merge with one of the humps that were born later.\\
(b) Phase diagram of the various traffic states that can occur close
to an on-ramp in the presence of small perturbations in the ramp flow.
We show the dependence of the traffic states
on the ramp flow and the ramp length for a flow of
1800 vehicles per hour and lane on the freeway. For small ramp flows,
free traffic (FT) survives. At higher inflows, two different kinds of
RH states can build up, either triggered stop and go waves (TSG) or 
oscillatory synchronized traffic (OST). High ramp flows are associated
with a homogeneous form of synchronized congested traffic (HST).}
\end{figure}
\clearpage
\begin{center}
{\large\bf Hyperlinks}
\end{center}
\begin{itemize}
\item PD Dr. Dirk Helbing:\\ 
http://www.theo2.physik.uni-stuttgart.de/helbing.html
\item Dr. Martin Treiber:\\
http://www.theo2.physik.uni-stuttgart.de/treiber.html
\item Dr. H. Y. Lee:\\
http://phya.snu.ac.kr/\~{ }agnes
\item Prof. Dr. D. Kim:\\
http://phya.snu.ac.kr/\~{ }dkim
\item Dr. Kai Nagel:\\
http://www.santafe.edu/\~{ }kai/
\item Prof. Dr. Michael Schreckenberg:\\
http://traffic.comphys.uni-duisburg.de/member/home\_schreck.html
\item Hot links to traffic research:\\
http://traffic.comphys.Uni-Duisburg.DE/hottrafficlinks.html
\item International Conference 
``Traffic and Granular Flow '99: SOCIAL, TRAFFIC, AND
GRANULAR DYNAMICS'':\\ 
http://www.theo2.physik.uni-stuttgart.de/tgf99/
\item Simulate traffic with a cellular automaton:\\
http://www.theo2.physik.uni-stuttgart.de/helbing/RoadApplet/
\item Preprint server cond-mat:\\
http://xxx.lanl.gov/find/cond-mat/
\end{itemize}
\end{document}